# Percolation-driven β-relaxation enables resonant acceleration of crystallization in amorphous phase-change materials

Yu-Yao Liu[#1], Liang Gao[#1], Jun-Ying Jiang[1], Yiming Zhou[2], Jan Luebben[3], Di Zhao[2], Xiaoling Lu[4], Maximilian J. Müller[2], Ulrich Boettger[3], Jiang-Jing Wang[*2,5], Hai-Bin Yu[*1] and Shuai Wei[*6,7]

[1]*Wuhan National High Magnetic Field Center and School of Physics, Huazhong University of Science and Technology, Wuhan, Hubei 430074, China*

[2]*Institute of Physics (IA), RWTH Aachen University, Aachen, 52056, Germany.*

[3]*Institut für Werkstoffe der Elektrotechnik II (IWE II), RWTH Aachen University, Aachen, 52056, Germany.*

[4]*Institut für Werkstoffe der Elektrotechnik I (IWEI), RWTH Aachen University, Aachen, 52056, Germany.*

[5]*Center for Alloy Innovation and Design (CAID), State Key Laboratory for Mechanical Behavior of Materials, Xi'an Jiaotong University, Xi'an, 710049, China*

[6]*Department of Chemistry, Aarhus University, 8000 Aarhus C, Denmark*

[7]*CENSEMAT Center for Sustainable Energy Materials, Aarhus University, Denmark.*

**ABSTRACT**:

**Amorphous phase-change materials enable fast and reversible switching in optical and electronic devices, yet crystallization kinetics are still controlled primarily through empirical thermal protocols. Here we identify a microscopic picture governing crystallization in the prototypical phase-change material $Ge_2Sb_2Te_5$, in which crystallization pathways are organized by the percolation of mobile atomic networks associated with β-relaxation. We show that this percolation transition distinguishes the dominance of diffusion-driven and diffusionless nucleation and growth during crystallization processes. We further demonstrate that frequency-selected ultrasonic excitation, applied in conjunction with heating, accelerates crystallization by enhancing percolation-mediated atomic dynamics. This acceleration is maximized near the β-relaxation frequency, consistent with resonant excitation of mobile atoms. Our results establish a direct link between glassy relaxation, atomic-scale percolation, and crystallization, and introduce a new route to modulating phase-change kinetics through targeted excitation of fundamental glassy dynamics.**

[#] These authors contributed equally.

[*]j.wang@xjtu.edu.cn (JJW), haibinyu@hust.edu.cn (HBY), shuai.wei@chem.au.dk (SW)

## Introduction

Amorphous materials represent a broad class of condensed matter systems characterized by the absence of long-range order, yet they exhibit rich structural and dynamical phenomena of both fundamental and technological importance[1, 2]. Their stability[3], relaxation dynamics[4, 5], and crystallization behavior[6] not only underpin our understanding of glass physics but also dictate their performance in practical applications, ranging from structural alloys to functional glasses[7, 8, 9, 10, 11]. One of the central challenges is to uncover how atomic-scale dynamics in amorphous states give rise to large-scale structural rearrangements that ultimately lead to crystallization.

Within this context, relaxation dynamics have emerged as a key theme[12, 13, 14, 15, 16, 17, 18]. The primary α-relaxation governs viscous flow and the glass transition but becomes essentially frozen below $T_g$ [19, 20, 21]. In contrast, the Johari-Goldstein β-relaxation -- arising from localized, fast atomic motions -- remains active well below $T_g$, providing a crucial source for atomic mobility[22, 23]. Recently, a double-percolation scenario of the two relaxation processes has been proposed[24] in metallic glasses, binary Kob-Andersen systems, and a metalloid Ni-P glass[24], in which α and β relaxations correspond to immobile and mobile particle percolations, respectively, whenever they are well separated. Despite the importance of percolation in glass dynamics, its relevance in covalent glasses remains unexplored, and the connection between the underlying atomic rearrangements and the pathways leading to long-range ordering during crystallization has yet to be established.

Amorphous chalcogenide phase-change materials (PCMs) (e.g. $Ge_2Sb_2Te_5$) form a distinct class of covalently bonded glasses, notable for their poor glass forming ability and fast crystallization at elevated temperature due to weakened covalency compared to conventional chalcogenide glasses[25] (e.g. GeSe). Their crystalline states also exhibit unusual properties absent in conventional crystalline chalcogenides, including high optical dielectric constants, large Born effective charge, and high Grüneisen parameters, an effective coordination number violating the 8-N rule[26]. They are one of the most promising candidates for non-volatile memory (NVM) and in-memory computing (IMC) applications[27, 28, 29]. Their capability of rapid phase transitions between amorphous and crystalline states, as well as the pronounced contrast in electrical resistance or optical reflectivity are utilized to encode digital information[30, 31, 32]. The practical performance of these devices, however, hinges critically on the crystallization kinetics of the amorphous phase[30, 33]. While fast crystallization enables rapid data writing, the ease of crystallization might undermine the stability of the amorphous state and data retention properties[34, 35, 36]. Thus, understanding and controlling the amorphous-to-crystalline transition dynamics is of fundamental and practical significance.

Recent studies have revealed the presence of β-relaxation in amorphous PCMs, manifesting as a low-temperature excess wing in mechanical loss spectra[37]. Crucially, the tunability of β-relaxation has been shown to strongly influence crystallization[38]. Despite these advances, a fundamental question remains unresolved: does the percolation scenario of mobile atoms, established in metallic and Kob–Andersen glasses, also govern β-relaxation in PCMs? If so, how does this percolation contribute to their unique crystallization kinetics? Addressing these questions is essential, given that PCMs differ significantly from metallic glasses in bonding and local structure—the former being often dominated by distorted defective octahedra with covalent bonding[39], and the latter by densely packed polyhedral motifs with nondirectional metallic bonds[40, 41]. Whether the same microscopic mechanism applies to PCMs remains elusive.

In this work, we focus on the prototypical PCM $Ge_2Sb_2Te_5$ (GST). Using molecular dynamics simulations based on a DeepMD neural network potential[39, 42], we demonstrate that β-relaxation in GST originates from the percolation of mobile atoms, which provides nucleation pathways for crystallization. Furthermore, we show both computationally and experimentally that ultrasonic excitation resonant with the characteristic frequency of β-relaxation can effectively modulate the percolation behavior, thereby accelerating crystallization. These findings not only elucidate the microscopic origin of β-relaxation in PCMs but also establish percolation as a key mechanism linking local dynamics to crystallization.

## Results
### The mobile atom percolation and β-relaxations

We performed the molecular dynamics simulations of dynamic mechanical spectroscopy (MD-DMS)[43, 44] with percolation analysis[24, 45] on the GST system. The details of the simulations are reported in the Methods of Supplementary Information (Sec.1.1). Figure 1a (upper panel) displays temperature-dependent loss modules ($E''$) with an oscillation period $t_\omega = 1$ ns (the corresponding frequency is 1 GHz), revealing an excess wing indicative of the β-relaxation at ~ 410 K. The MD-DMS result well reproduces the excess wing observed in a previous experimental study of GST[37].

The percolation analysis is based on two critical thresholds: the displacement threshold $u_c$ and the distance threshold $r_c$ (Supplementary Figure 1). The value of $u_c$, determined from the displacement $u$ of atoms over one MD-DMS cycle at the maximum of the non-Gaussian parameter[41, 46, 47, 48] (Supplementary Figure 2), specifies whether an atom is classified as mobile ($u > u_c$) after this cycle. The $r_c$, extracted from the first minimum of the pair distribution function $g(r)$, defines the connectivity criterion for grouping atoms into clusters: when the distances between two atoms is smaller than $r_c$, they belong to the same cluster.

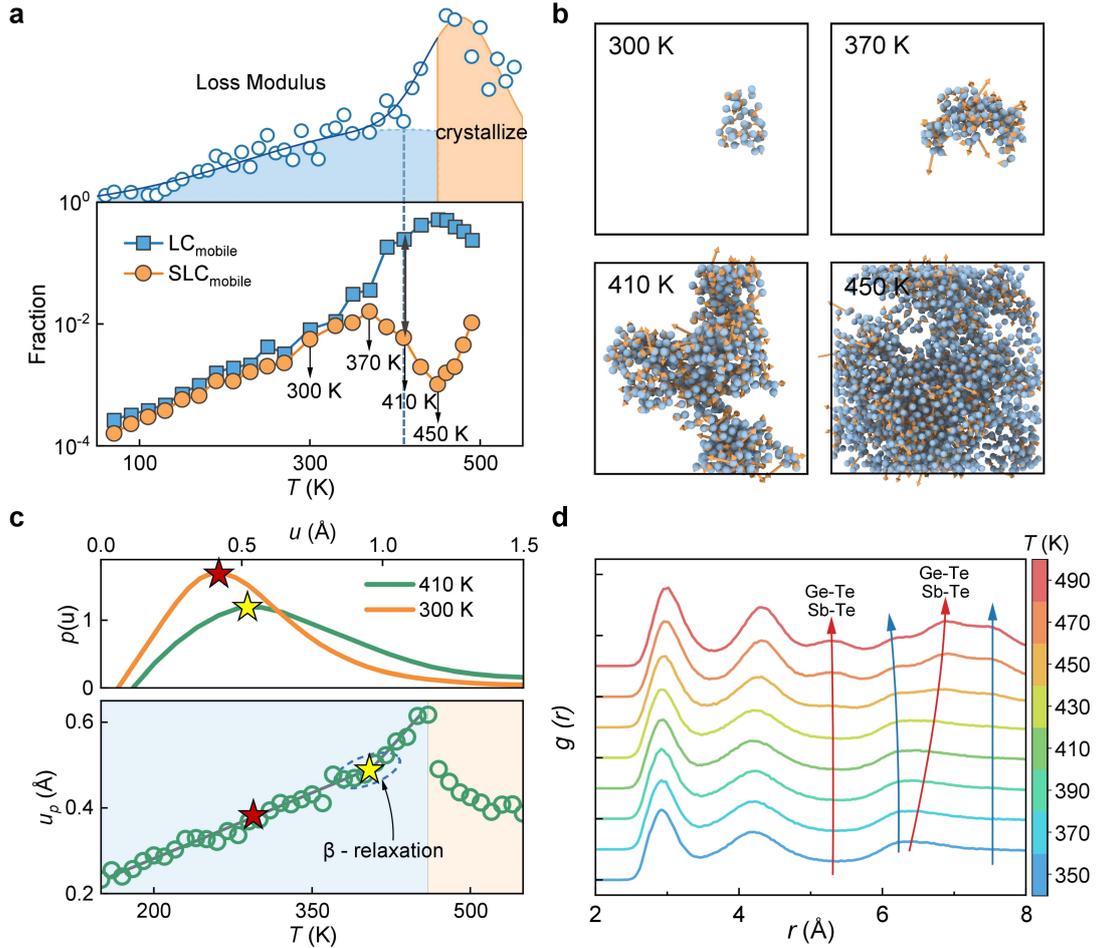

**FIG.1 Correlation between mobile percolation and β relaxation. a** The upper panel shows the loss modulus $E''$ of GST model with an oscillation period $t_\omega = 1$ ns. The lower panel presents the sizes of the largest cluster (LC, blue) and the second-largest cluster (SLC, orange) as a function of temperature. The double-headed arrow indicates the percolation. **b** Configurations of the largest cluster at $T = 300$ K, $T = 370$ K, $T = 410$ K and $T = 450$ K. **c** The upper panel shows the displacement distributions $p(u)$ in MD-DMS at 300 K and 410 K, while the lower panel shows the evolution of its peak with temperature. **d** Pair distribution function $g(r)$ evaluation during heating process, showing the crystallization near 450 K. The red arrows indicate Ge-Te and Sb-Te atom pairs that appear after 450 K. The blue arrows indicate the other pairs.

As shown in the lower panel of Fig. 1a, we track the evolution of the two major mobile clusters, the largest (LC) and the second largest (SLC) with $u_c = 0.9$ Å and $r_c = 3.5$ Å (see Supplementary Figure 3). Both clusters exhibit similar growth trends at lower temperatures but begin to diverge significantly above 370 K. Beyond this divergent temperature, the LC continues to grow while the SLC decreases correspondingly, indicating the dominance of the LC through merging other clusters, and marking the onset of percolation. This percolation temperature coincides with the β relaxation in upper panel, indicating that mobile atom percolation underpins the β-relaxation in the PCM despite complex interactions. To provide an intuitive illustration, Fig.1b shows LC at 300 K, 370 K, 410 K and 450 K, highlighting the evolution of mobile atom clusters. We observe that the LC grows significantly from 370 K to 410 K, spans the system at 410 K, and gradually takes over the entire configuration space thereafter.

The upper panel of Fig. 1c shows the atomic displacement distribution $p(u)$, which shifts toward larger $u$ values at 410 K compared to 300 K. And the lower panel of Fig. 1c plots the peak $u_p$ of $p(u)$ as a function of MD-DMS temperature. We find that $u_p$ increase drastically when $T$ approaches the β-relaxation (yellow star), indicating an increased atomic mobility. However, this growth is interrupted around 450 K, where it turns into a decline. This decline in $u_p$ coincides with a marked structural change: as shown in the Fig. 1d, the pair distribution function $g(r)$ begins to split into multiple distinct peaks above 450 K, indicating the onset and progression of crystallization during the MD-DMS process.

**The relationship between the crystallization and percolation**

Aiming to unravel the crystallization dynamics of the GST system and their connection to β relaxation, we employ a structure genetic algorithm, cluster alignment strategy (see Methods), to determine the crystalline atoms. In this strategy, reference templates for each atomic species are first derived from the crystallized GST structure using the *pair* mode, with the extracted templates illustrated in the inset of Fig. 2a. These templates serve as structural benchmarks for identifying crystalline atoms in *template* mode. Figure 2a shows the fraction of crystal as a function of temperature, using an alignment score of 0.08 as the criterion for crystal identification. The alignment score quantifies the structural similarity between an atomic cluster and its template, with smaller values indicating a higher degree of crystalline order. A sharp increase in the fraction is observed shortly after the percolation of mobile atoms at approximately 410 K, highlighting a strong connection between β relaxation and crystallization. This correlation is further illustrated in Fig. 2b.

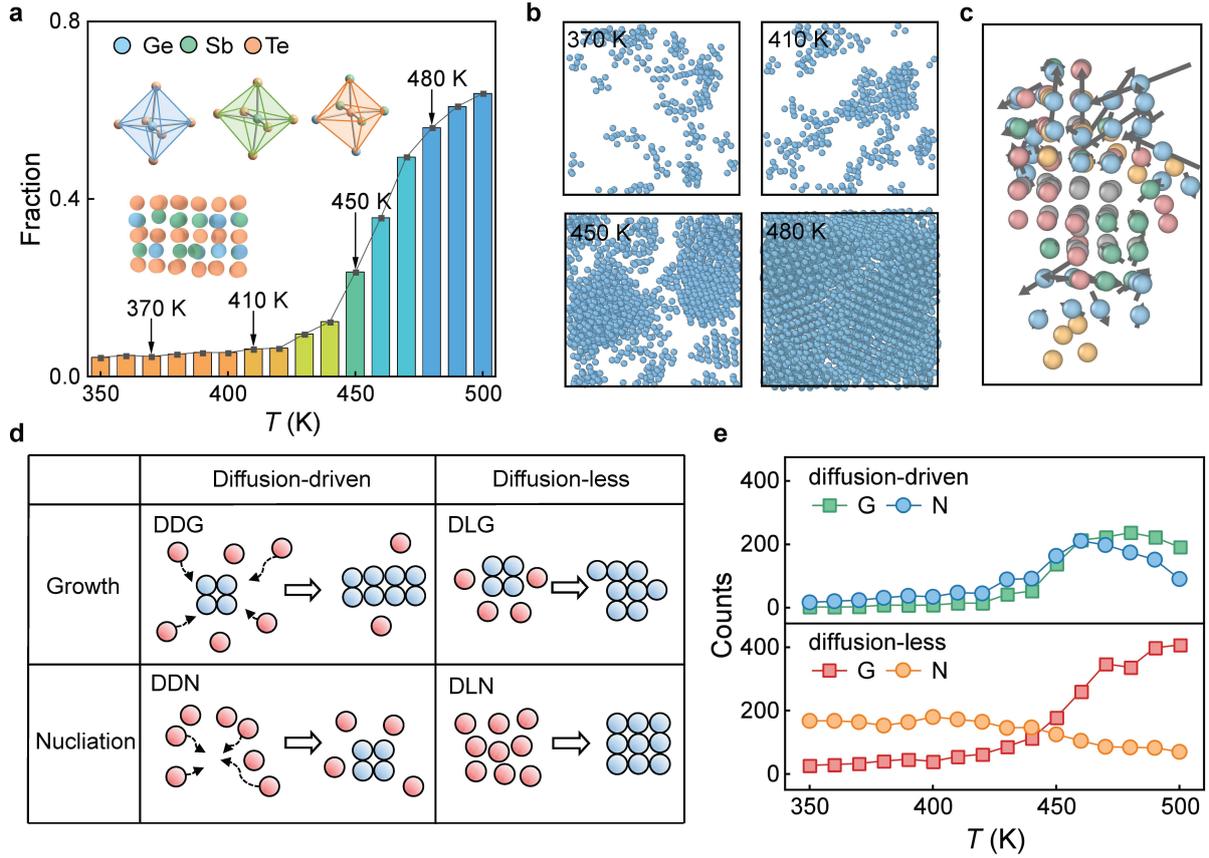

**FIG. 2 Crystallization pathways near the β-relaxation. a** Crystallized atoms fraction as a function of temperature from the MD-DMS. The insets show the crystalline structure and three different atomic coordinates contained in GST crystal. **b** The crystalline atomic configurations at 370 K, 410 K, 450 K, and 480 K. **c** A typical crystallization process observed during MD-DMS. Red, green, orange and blue represent different pathways of crystallization. The arrows mean the direction of atomic displacement. **d** Schematic illustration of distinct crystallization pathways. Blue regions indicate crystalline domains, while red regions correspond to amorphous structures. **e** The counts of distinct crystallization pathways during MD-DMS at each temperature.

As shown in Fig. 2c, a representative crystallization event at 450 K exhibits a complex interplay of mechanisms, including both diffusion-driven and diffusion-less crystallization. These processes are systematically categorized into four pathways, as illustrated in Fig. 2d. The classification procedure is as follows. Crystalline atoms already present in the reference configuration are labeled as old crystalline atoms, $\{r_i\}_{old}$. After one MD-DMS cycle, atoms that newly become crystalline are labeled as new crystalline atoms, $\{r_i\}_{new}$. Among these, atoms with displacements greater than $u_c$ (as in the percolation analysis) are classified as diffusion-driven crystallization. Some of these atoms migrate to the vicinity of existing crystalline atoms and grow into new crystalline atoms (diffusion-driven growth, DDG), while others diffuse and nucleate independently (diffusion-driven nucleation, DDN). In contrast, new crystalline atoms with displacements smaller than $u_c$ are treated as diffusive-less crystallization. Those found within a cutoff radius $r_c$ of the old crystalline atoms are attributed to growth (diffusion-less growth, DLG), while those beyond $r_c$ are regarded as arising from nucleation (diffusion-less nucleation, DLN).

Figure 2e shows the distribution of newly crystallized atoms classified by their pathways. At low temperatures (below 410 K, consistent with β relaxation), crystallization is primarily driven by diffusion-less nucleation (DLN), which facilitates the initial accumulation of crystal nuclei. As the temperature rises above 410 K, both diffusion-driven crystallization (DDG and DDN) and diffusion-less growth (DLG) pathways exhibit a pronounced increase. This mechanistic transition aligns with the peak in crystallization rate observed in Fig. 2a, highlighting the key role of diffusion in accelerating crystallization following the β-relaxation. All data are averaged over ten independent DMS cycles at each temperature to ensure statistical robustness (Fig. S4).

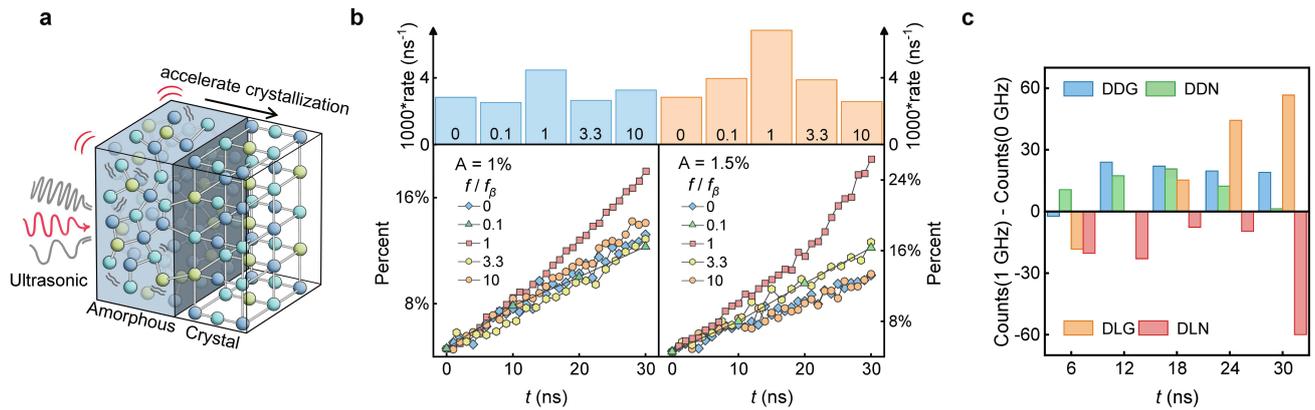

**FIG.3 Ultrasonic accelerated crystallization. a** Schematic of the ultrasonic-accelerated crystallization mechanism. **b** Crystalline fraction and crystallization rate versus time at ultrasonic amplitudes of 1% and 1.5% at 410 K for frequencies $f$ with respect to the intrinsic frequency $f_\beta$ = 1 GHz of the β relaxation at 410K. **c** Ultrasonic-treated versus untreated systems: $\Delta$Counts($t$) = Counts(1 GHz, $t$) - Counts(0 GHz, $t$). Time evolution of different crystallization pathways over 30 ns at 410 K. DDG: diffusion-driven growth, DDN: diffusion-driven nucleation, DLG: diffusion-less growth and DLN: diffusion-less nucleation.

**Ultrasonic resonance accelerated crystallization**

Building on our findings linking crystallization dynamics to the atomic diffusion, we next investigate whether this diffusion could be externally modulated to achieve controllable crystallization. For this purpose, we apply ultrasonic excitation to our simulated GST system. Figure 3a schematically illustrates the mechanism of ultrasonic-assisted crystallization. Ultrasonic vibrations with amplitudes ranging from 0.5 % to 1.5% (relative to the simulation box size) with frequencies between 0.1 and 10 GHz are applied at the β relaxation temperature $T_\beta = 410$ K. Notably, as shown in Fig.3b, the optimal enhancement in crystallization fraction and rate occurs at 1 GHz, which coincides with the intrinsic frequency 1 GHz of the β relaxation at this temperature, highlighting a resonant coupling between mechanical modulation and atomic mobility. In addition, a higher ultrasonic amplitude shows to larger effects on accelerating crystallization at the resonant frequency.

To elucidate the mechanism underlying ultrasonic-accelerated crystallization in GST, we analyze the crystallization pathways during ultrasonic process, the same as previously classified. For comparison, the same analysis is performed on the untreated system. The crystallization pathway distribution (Fig.3c) reveals that the ultrasonic stimulation fundamentally modifies the dominant crystallization pathway. The diffusion-less nucleation dominates crystallization in the absence of ultrasonic stimulation. However, under 1 GHz ultrasonic vibrations, the contribution of DLN crystallization decreases significantly, while DDG and DDN are markedly enhanced within 30 ns. The pronounced increase of DLG after 15 ns is attributed to the accumulation of numerous nuclei formed via DDN within the initial 15 ns under ultrasonic stimulation, thereby promoting a transition of the dominant mechanism from nucleation to growth. This is further supported by the observed decrease in DDN after 15 ns, coinciding with the increase in DLG. It is worth noting that Fig. 3c shows the instantaneous change in the number of new crystals, while Fig. 3b shows the cumulative effect over 30 ns. Extended-timescale and other amplitudes statistics are provided in the Supplementary Information.

**Experimental validation**

Next, we experimentally validate the theoretical concept by examining laser-induced crystallization in an amorphous GST thin film under ultrasonic vibrations. Note that the simulations were performed at a constant temperature ($T = 410$ K) and frequency ($f = 0, 0.1, 1, 3.3$ and 10 GHz). These parameters were chosen for computational feasibility, allowing us to observe crystallizations and β-relaxations within reasonable computational timescales. In experiments, replicating these exact conditions is unrealistic due to sample size effects, heating dynamics, and the limitations of temporal resolutions in experimental setups. A laser heating would continuously and rapidly ramp up temperature, thus increasing the frequency of the β relaxation. As such, the resonance frequency under such a condition is unknown, but must be higher than the experimentally reported 1 Hz in our previous PMS experiment[37]. Considering the prior studies on bulk metallic glasses showing that the ultrasonic excitation in the megahertz range accelerates crystallization[49, 50], we employed a commercial ultrasonic device generating ultrasonic waves between 5 and 35 MHz for our "pump-probe" laser set-up. A pump laser with a pulse width $t$ is used to heat the sample film and a probe laser monitors the film reflectivity $R$ changes (see Methods).

The crystallization behaviors of GST films can be characterized by measuring the power-time-effect (PTE) diagrams under ultrasonic vibrations at the frequency 5 MHz and its overtone ($n$= 3, 5, 7) frequencies 15, 25, and 35 MHz (Fig. 4a). The top panel corresponds to the reference condition (0 MHz). With increasing pulse width $t$, the reflectivity $R$ shows a sharp increase at around 200-400 ns for the laser power $P$ ranging from 5 to 20 mW, stemming from the onset of crystallizations. The onset time does not show a clear dependency of frequencies. However, a close examination of the reflectivity colormap reveals a higher reflectivity region (darker red color) on a longer timescale (e.g. $10^4$ ns) at 5 MHz. To better illustrate the difference, we show the differential PTE by subtracting the reference PTE (0 MHz) (see Fig. 4b). Clearly, the differential PTE at 5 MHz shows a reflectivity increase, although no such a strong increase is observed at other frequencies. This suggests a higher crystalline fraction at 5 MHz, resulting from faster crystallization kinetics. Since the onset time of crystallization has not been altered, we can infer that the crystal growth rate plays a dominant role in determining the observed crystal fractions.

For a higher time- and power- resolution, we conducted the fine-step PTE measurements for 5 and 0 MHz. The result reproduces the reflectivity difference between 5 and 0 MHz (see Supplementary Fig.21). Figure 4c shows the film reflectivity contrast, $C = (R_{After}- R_{Before})/R_{Before}$, before and after the pump laser. The difference between 5 and 0 MHz is evident when the contrast increases above 42%. Figure 4d shows the pulse width $t$ required to achieve contrasts of 42%, 45%, and 47% at powers ranging from 14.2 to 18.4 mW. At 5 MHz, the pulse width decreases by a factor of 2 – 9 for 45% contrast and by a factor of 2 – 4 for 47% contrast, indicating a substantially shorter time to reach a given crystalline fraction.

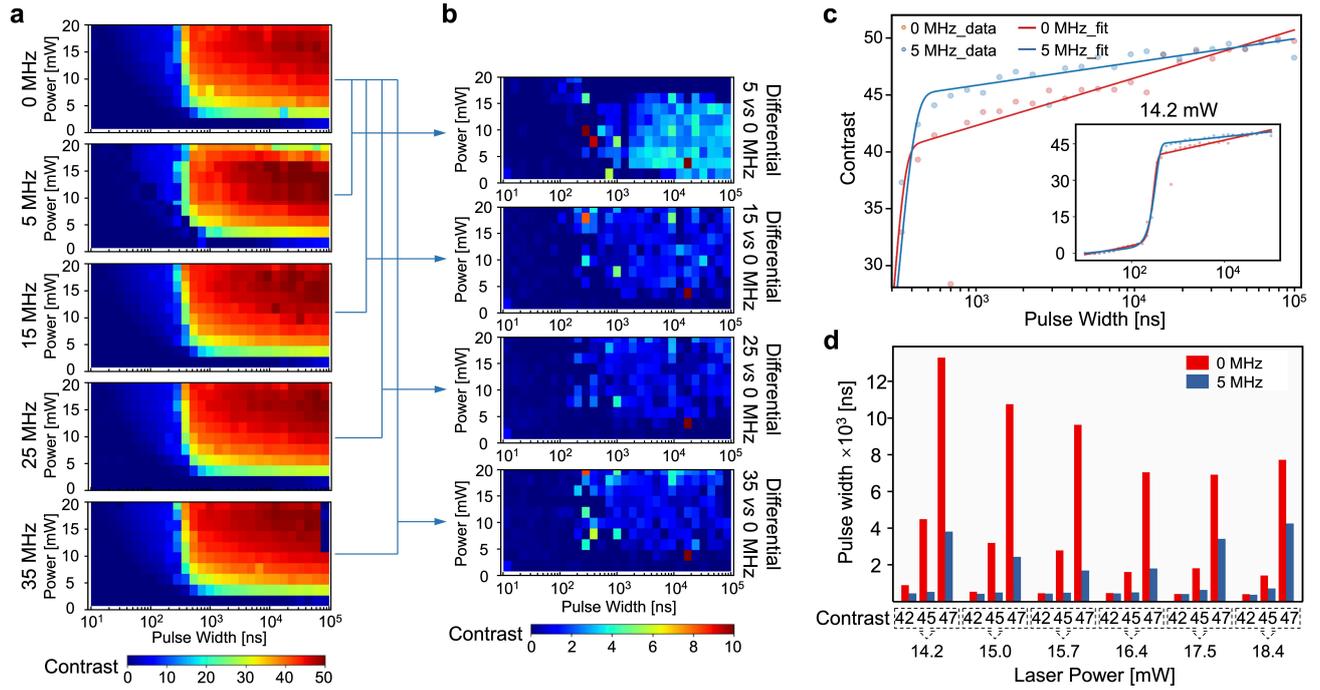

**FIG. 4 Crystallization under ultrasonic vibrations. a** The power-time-effect (PTE) diagram for crystallization at various ultrasonic wave frequencies of 0, 5, 15, 25, 35 MHz. The color map indicates the reflectivity change for a given pulse width and a pulse power. The regions with green to red colors indicate a higher reflectivity and thus a higher volume fraction of crystals. **b** The differential PTE diagram shows a high reflectivity region for 5 MHz (top panel), suggesting faster crystallization kinetics. **c** Comparison of contrast in fine-step scans for 0 and 5 MHz at a representative power 14.2 mW (see Supplementary Fig.22 for other powers). **d** Laser pulse width required to reach the certain contrast of 42%, 45% and 47% for the given power.

## Discussion and outlook

The MD simulations demonstrated that mobile-atom percolation underlies β-relaxation in covalently bonded amorphous GST, organizes crystallization pathways, and enables frequency-selective acceleration of crystallization. Despite distinct bonding types, the percolation-β relaxation relationship has been found for metallic glasses, Kob-Andersen model, and a Ni-P glass[24, 51]. In covalent PCMs, β-relaxations manifest as an excess wing of the α-relaxation[37], whereas in metallic glasses they take diverse forms, including peaks (e.g., La-based alloys), shoulders, or wings[12]. In the Kob–Andersen mixture, they appear as a wing, while in the metalloid alloy $Ni_{80}P_{20}$, they can manifest as either a shoulder or a wing depending on probe frequency[24, 51]. These observations suggest that, compared with metallic glasses displaying shoulders or peaks, PCMs possess a relatively narrower distribution of energy barriers for flow events (i.e., smaller dynamic heterogeneity). Within the double-percolation scenario[24], this corresponds to mobile- and immobile-particle percolation temperatures lying relatively close to each other. This may also account for the vanishingly small β-relaxations observed in conventional covalent glasses such as GeSe, GeSe$_2$, and Ge$_{15}$Te$_{85}$[37], where stronger covalency with a more rigid network structure produce more homogeneous dynamics and an even narrower barrier distribution. While percolation concept has been shown important for glass formation[52], fragility[53], conductivities[54], vibrational properties[55] in various types of glasses, our results identify it as a microscopic origin of β-relaxations in PCMs and directly link it to crystallization behavior.

The ultrasonic-accelerated crystallization in thin GST films is qualitatively in agreement with the MD simulations, supporting a percolation-mediated mechanism in which frequency-selective ultrasonic resonance facilitates crystallizations. As shown in simulations, the resonance accelerates both nucleation and growth (Fig. 3c) by providing local atomic mobility through percolation pathway. In the experiment, it is likely that there are more nuclei formed under ultrasonic resonance; however, a higher concentration of nuclei may not be visible in contrast measurements due to the limited reflectivity resolution. Nevertheless, an increased number of nuclei facilitates the growth process via DDG and DLG, as shown in simulations (Fig.3c). The effect of a higher growth rate under ultrasonic resonance becomes increasingly apparent when crystals grow to a larger volume fraction (Fig.3b and 4b). Both the simulation and experiment for GST are reminiscent of an earlier report of Pd-based bulk metallic glasses[50, 56], where an isothermal crystallization (at slightly above $T_g$) completed up to 10 times faster under ultrasonic vibrations at 1.2 MHz and were phenomenologically attributed to resonance with β-relaxations. In contrast to those studies, our work establishes a microscopic percolation-based mechanism linking β-relaxation to crystallization pathways and demonstrates how frequency-selective excitation can be used to deliberately accelerate crystallization in PCMs. More importantly, while accelerated crystallization is generally detrimental in metallic glasses, because it often leads to degraded mechanical properties, it is a desired and enabling functionality in PCMs, where fast and controllable crystallization underpins device operation.

Earlier work reported the suppression of β-relaxations slowing down crystallization, for instance, through annealing[38] and constrained sample geometry[34]. The present study demonstrated that ultrasonic resonance as an approach to enhance β-relaxations to accelerate crystallization. The present accessible frequencies of our ultrasonic devices are constrained by overtone modes, which limits the ability to systematically explore the full frequency response. Probing additional frequencies in the vicinity of 5 MHz would be informative for a more detailed mapping of the frequency dependence of crystallization kinetics. Moreover, extending such studies to different forms of acoustic excitation (e.g., longitudinal versus shear waves, or continuous versus pulsed modes) would further elucidate the role of specific vibrational modes in modulating atomic mobility and crystallization behavior. Ultrasonic-modulated dynamics may, in principle, offer a route for designing the better performance of PCM-based switchable devices. For instance, ultrasonic waves and electrical pulses can be synchronized and applied to the memory cells, while operating the SET-switching for a faster switching. For the period of data retention, ultrasonic wave is held off to ensure an un-deteriorated amorphous stability. The experiment in this work is based on optical switching, whereas electronic switching is expected to exhibit similar ultrasonic-accelerated crystallization. Finally, gaining a more profound understanding of the interplay between external perturbation, intrinsic relaxation dynamics, percolation, and crystallization, is intriguing and calls for future experimental and computational studies at the atomic level in PCMs.

## Methods

### Molecular dynamics simulations and dynamic mechanical spectroscopy

All molecular dynamics (MD) simulations were simulated using the LAMMPS package[57]. Our model system contains $N = 6912$ atoms of the prototypical phase-change material $Ge_2Sb_2Te_5$. This system interacts with a DeepMD neural network potential developed by O. Abou El Kheir et al[39]. The glassy state was achieved through quenching from the liquid state at 1000 K to 10 K with a cooling rate of $10^{12}$ K/s in the NVT ensemble. Then the glassy system was heated to near its crystallization temperature at a rate of $10^{10}$ K/s for subsequent calculations.

In selected temperature ranges during the heating, the model was subjected to a sinusoidal shear strain with an oscillation period $t_\omega$ and a strain amplitude $\varepsilon_A$, along the $x$ direction of the model. And the strain was expression as $\varepsilon(t) = \varepsilon_A \sin(2\pi t / t_\omega)$. This can result in stress $\sigma(t)$ and the phase difference $\delta$ between stress. The stress was measured and described by $\sigma(t) = \sigma_0 + \sigma_A \sin(2\pi t / t_w + \delta)$. The shear storage and loss moduli were calculated according to $E' = \frac{\sigma_A}{\varepsilon_A} \cos(\delta)$ and $E'' = \frac{\sigma_A}{\varepsilon_A} \sin(\delta)$, respectively.

### Cluster and percolation analysis

At a given time interval $\Delta t$, mobile atoms are defined by the van Hove function $p(u, \Delta t)$. The van Hove function $p(u) \equiv [P(u + \Delta u) - P(u)]/\Delta u$ is used in which $P(u)$ is the cumulative distribution quantifying the probability of finding the value $X \leq u$. The one cycle of MD-DMS is selected as the time interval $\Delta t$. The $u_c$ is determined by the displacement of the mean square displacement corresponding to the peak of Non-Gaussian Processes (NGP, Supplementary Fig. 2). NGP characterizes the inhomogeneity of dynamics. Mobile atoms have displacement greater than $u_c$.

According to the radial distribution function $g(r)$, atoms that are close to each other are defined as the same cluster. When the distance between atoms is below that marking the first minimum of $g(r)$ which is $r_c$, atoms by definition belong to the same cluster. According to supplementary Fig.1, the value of $r_c$ is different for crystal and amorphous. For amorphous GST, $r_c = 3.5$ Å. And for crystal GST, $r_c = 3.8$ Å. Those with a distance of less than 3.8 Å from the old crystal are defined as crystallized by growth.

A sequence of clusters with varying sizes (the number of atoms) can be identified based on $u_c$ and $r_c$. The largest cluster is denoted as LC, the second largest cluster as SLC. The criterion applied to determine the percolation temperatures is that the size of LC is 100 times that of SLC.

**Cluster alignment for crystalline cluster identification**

To accurately identify crystalline clusters within mixed disordered and ordered GST structures, we employed a two-step cluster alignment strategy.

1. Template Extraction: Highly crystallized GST configurations were first analyzed to extract representative local structures. Using the pair alignment mode[58, 59], clusters were compared with an alignment score of 0.06, and the most frequently occurring clusters were selected as templates. These templates capture the dominant local motifs within the crystalline regions and serve as references for subsequent identification.

2. Crystalline Cluster Identification: The extracted templates were then applied to mixed GST structures using the template alignment mode. Clusters exhibiting an alignment score below 0.08 were classified as crystalline. To ensure structural fidelity, each identified cluster was allowed to contain no more than two vacancies. This procedure enables precise discrimination of crystalline motifs within heterogeneous environments, facilitating quantitative analysis of crystallization in GST systems.

**Laser induced crystallization under ultrasonic excitation**

A "pump-probe" laser set-up is used to measure the PTE diagrams of a phase-change thin film device (Supplementary Figure 20). The wavelength of the pump and probe lasers are 658 nm and 639 nm, respectively.

The pump laser shines a continuous pulse to locally heat the thin film and trigger crystallization, while the probe laser is used to measure the reflectance changes. The thin film device contains a commercial Quartz Crystal Microbalance (QCM) substrate (openQCM-Q1 by Novatech S.r.l., Italy), for generating ultrasonic vibrations at the fundamental and overtone frequencies. The substrate is deposited with a thin GST film (100 nm in thickness) sandwiched by a 10 nm thermal barrier as well as a 100 nm capping layer of $(ZnS)_{80}(SiO_2)_{20}$ to prevent the heat loss and potential evaporation. The frequencies and dissipation during the measurements were recorded and reported in the Supplementary Table 1. The details of laser and ultrasonic set-up are described in Supplementary Methods Sec.1.2.

## Data availability

All of the data used to generate the figures are available at

## Code availability

The simulation package LAMMPS (https://www.lammps.org) was used for all molecular dynamics simulations. The code and scripts of the analysis are available at

## Acknowledgements

The computational work was carried out on the public computing service platform provided by the Network and Computing Center of HUST. We thank for the support from the National Science Foundation of China (NSF 52071147). Discussions with Baoshuang Shang are appreciated. Authors thank Marco Bernasconi for helpful comments and providing the ML potential and the helpful discussion and support from Matthias Wuttig. J.W. thanks the support of XJTU for the work at CAID. S.W. thanks Villum Fonden (42116) for financial support. AI tools were only used for stylish polishing and grammar corrections, which are checked and verified by authors.

## Author contributions

H.B.Y. and S.W. conceived the research. Y.Y.L. and L.G. conducted the simulations supervised by H.B.Y.. J.J.W, Y.Z., S.W., J.L., D.Z., X.L., M.J.M., U.B. designed and performed the experiments. Y.Y.L, H.B.Y., J.J.W and S.W. wrote the manuscript with input from other authors.

## Competing interests

SW and XL are inventors of a patent EP3970147 and 11817146 (US). The other authors declare no competing interests.

## Additional information

Supplementary information
The online version contains supplementary material available at